\documentclass[]{CHRONOS_s}  
\usepackage{amsmath,amsfonts,amssymb}
\usepackage{graphicx}
\usepackage{setspace}
\usepackage{tocloft}
\usepackage{comment}

\author{Daiki Tanabe\supit{a,b,c,\dag}, Masaya Hasegawa\supit{c,d}, Masashi Hazumi\supit{a,c,d,e,f}, Nobuhiko Katayama\supit{e}, Shuhei Kikuchi\supit{g}, Adrian Lee\supit{h,i}, Haruki Nishino\supit{j}, Satoru Takakura\supit{k}
\skiplinehalf
\supit{a}Department of Particle and Nuclear Physics, School of High Energy Accelerator Science, The Graduate University for Advanced Studies (SOKENDAI), Tsukuba, Ibaraki 305-0801, Japan; \\
\supit{b}Center for High Energy and High Field Physics, National Central University, Taoyuan 32001, Taiwan;\\
\supit{c}Institute of Particle and Nuclear Studies (IPNS), High Energy Accelerator Research Organization (KEK), Tsukuba, Ibaraki 305-0801, Japan;\\
\supit{d}International Center for Quantum-field Measurement Systems for Studies of the Universe and Particles (QUP), High Energy Accelerator Research Organization (KEK), Tsukuba, Ibaraki 305-0801, Japan;\\
\supit{e}Kavli Institute for the Physics and Mathematics of the Universe (Kavli IPMU, WPI), UTIAS, The University of Tokyo, Kashiwa,
Chiba 277-8583, Japan;\\
\supit{f}Japan Aerospace Exploration Agency (JAXA), Institute of Space and Astronautical Science (ISAS), Sagamihara, Kanagawa 252-5210, Japan;\\
\supit{g}Yokohama National University, Yokohama, Kanagawa 240-8501, Japan;\\
\supit{h}Department of Physics, University of California, Berkeley, CA 94720, USA;\\
\supit{i}Physics Division, Lawrence Berkeley National Laboratory, Berkeley, CA 94720, USA;\\
\supit{j}Research Center for the Early Universe, School of Science, The University of Tokyo, Tokyo, Japan;\footnote{ \ Present affiliation: Japan Synchrotron Radiation Research Institute (JASRI), 1-1-1, Kouto, Sayo-cho, Sayo-gun, Hyogo 679-5198, Japan}\\
\supit{k}CASA \& Department of Astrophysical \& Planetary Sciences, University of Colorado Boulder, Boulder, CO, 80309, USA}

\authorinfo{\dag Daiki Tanabe, dtanabe@post.kek.jp}

\title{High-precision temperature monitoring system for room-temperature equipment in astrophysical observations}

\cftpagenumbersoff{figure}
\cftpagenumbersoff{table} 
\begin{document} 
\maketitle

\begin{abstract}
We present a precise thermometry system to monitor room-temperature components of a telescope for radio-astronomy such as cosmic microwave background (CMB) observation. The system realizes precision of 1 mK${\rm \sqrt{s}}$ on a timescale of 20 seconds at 300 K. We achieved this high precision by tracking only relative fluctuation and combining thermistors with a low-noise measurement device. We show the required precision of temperature monitors for CMB observation and introduce the performance of our thermometry system. This precise room-temperature monitoring system enables us to reduce the low-frequency noise in a wide range of radio-astronomical detector signals observation and to operate a large detector array perform at its designed high sensitivity.
\end{abstract}


\keywords{temperature monitor, thermistor, readout, mirror, cosmic microwave background}


\begin{spacing}{2}   

\section{Introduction}
\label{sect:intro}
Large detector arrays are now commonly used in astronomical and cosmological observations. For millimeter and sub-millimeter imaging, the transition-edge sensor (TES) bolometer allows high-precision measurements of radiation power. The sensitivity of a single TES bolometer approaches the photon noise limit. The multi-pixel TES bolometer array is thus becoming the standard for improving sensitivity and expanding the field of view.

One of the major applications of the TES bolometer array is measuring the cosmic microwave background (CMB). The odd-parity polarization pattern of the CMB, which is called the {\it B}-mode, provides us with a probe of the inflation hypothesis in cosmology on large angular scales~\cite{seljak_inflation}. Testing inflation with the {\it B}-mode is, however, a challenge owing to the faint amplitude of the {\it B}-mode on the degree scale. It is lower than the temperature anisotropy by five orders of magnitude.

A standard strategy for measuring the {\it B}-mode precisely is improving statistical sensitivity by increasing the number of detectors. Several ground-based CMB experiments use $O(10,000)$ TES bolometers with designed array sensitivities reaching $O(1) \ {\rm \mu K \sqrt{s}}$~\cite{PBSA,AdvACT_detector,BICEP3_2014,SPT3G_TES}. For example, each receiver of the Simons Array (the POLARBEAR-2 telescope system) has 7,588 TES bolometers. Furthermore, the number of detectors for next-generation CMB experiments is projected to be $O(500,000)$ in the late 2020s~\cite{CMBS4_whitepaper}. 

Despite the low noise of a single detector and the large number of detectors in an array, noise correlated along multiple detectors cannot be mitigated by averaging signals over the array. One of the sources of correlated noise is 
the temperature fluctuation of the instruments. For example, the temperature fluctuation of the focal plane, where the detectors are 
placed, modulates the detector responsivity.

For a large array with more than $O(10,000)$ detectors, the temperature fluctuation of instruments outside the cryostat is also not negligible compared with the statistical sensitivity. There have been studies on the pointing error caused by the structure deformation of a mirror and a telescope due to the temperature variation~\cite{SPT_overview, ACT_optics, ACTpol_instrument}. The temperature fluctuation of the mirror also modulates the polarizations of the reflection and the emission from the mirror surface. This effect is seen by all detectors optically coupled with the given area of the mirror. Another example is the temperature fluctuation of readout electronics which changes the circuit gain. QUIET experiment deployed temperature regulation for readout electronics~\cite{QUIET_2013}. In recent experiments with large detector arrays, the fluctuation in readout performance is correlated among detectors owing to multiplexing. These correlated noises cannot be suppressed by averaging over detectors.

Whether we regulate temperature or correct the fluctuation effect in the offline analysis, we have to monitor the temperature with sufficient precision. If we observe a large angular scale with a relatively long duration, it is necessary to track temperature fluctuation during observations. There has been no report about room-temperature thermometry for noise reduction from the third-generation CMB experiments. Several radio-astronomical experiments have measured the room-temperature area for a telescope foundation or mirror deformation, but the precision was insufficient for reducing the noise of the CMB experiment~\cite{ALMA_underground,BIMA_temperature,6mBTA_temperature}.

In this paper, we show the requirements for precision of room-temperature monitoring system and demonstrate that our thermometry system satisfies these requirements. We focus on the mirror and the readout electronics as room-temperature instruments on the basis of the design of the POLARBEAR-2 telescope system in the Simons Array. We assume the specifications of the first telescope deployed in 2019~\cite{Kaneko_deploymentPB2A}.

In section 2, we discuss the temperature coefficients between instrumental temperature and a signal using the parameters of POLARBEAR-2, and then derive the requirements for the temperature monitoring system. Section 3 introduces the design concept and components of the temperature monitoring system. In section 4, we summarize the monitor design and discuss the source of systematic uncertainty in temperature measurement.

\section{Requirements for the precise temperature monitoring}
\label{sect:requirement}
\subsection{General formula for the temperature coefficient}
To avoid degradation of the designed performance of a detector array, an instrumental temperature fluctuation should be regulated or corrected so that the excess noise stays below the statistical sensitivity of the detector array. As we show later, the amplitude of the temperature fluctuation corresponding to a statistical sensitivity of $O(10000)$-TES array is on the order of $\mathrm{mK\sqrt{s}}$. 

First we express the timestream of the detector signal as 
\begin{equation}
d(t)=s(t)+n(t)+\sum_i a_i T_i(t),
\end{equation}
where $s(t)$, $n(t)$, and $T_i(t)$ are functions of time indicating the signal, noise, and temperature of the $i$-th instrument. Each instrumental component has a temperature coefficient, $a_i$, which relates temperature to the detector signal. It is a first-order perturbation expansion of the temperature dependence. When the temperature fluctuation is correlated among the detectors, we can treat $d(t)$ as an averaged timestream over an array.

The measured temperature $T'_i(t)$ is
\begin{equation}
T'_i(t)=T_i(t)+\delta T_i(t),
\end{equation}
where $\delta T_i(t)$ is the uncertainty in temperature measurement. By subtracting this measured temperature from the detector signal, we correct the detector noise caused by temperature fluctuation. This operation can be expressed via
\begin{equation}
\begin{array}{lll}
d'(t)&\equiv& d(t)-\sum_i a_i T'_i(t)\\
&=&s(t)+n(t)-\sum_i a_i \delta T_i(t).
\end{array}
\end{equation}
Thus, to avoid inducing excess noise, we require the uncertainty in temperature measurement to be below the original array noise translated via the temperature coefficient as
\begin{equation}
\label{eq:NSD}
{\rm NSD}[\delta T_i]< \left| \frac{{\rm NSD}[n]}{a_i} \right|,
\end{equation}
where ${\rm NSD}[\delta T_i]$ and ${\rm NSD}[n]$ are the noise spectrum densities of $\delta T_i(t)$ and $n(t)$, respectively. We focus on the relative temperature fluctuation because the CMB observations do not involve offset, but only measure the spatial variation of the CMB. Therefore, we evaluate the detector array noise, uncertainty in thermometry, and actual temperature fluctuation in terms of the noise spectrum density.

In the case of performing a temperature regulation, the $\delta T_i$ is replaced with the temperature fluctuation after the regulation. In conclusion, all instrumental temperature fluctuations have to be suppressed to stay below the right side term of Eq.~(\ref{eq:NSD}) by regulation or correction.

In the case of POLARBEAR-2, the designed array sensitivity of one receiver to polarization is $4.1 \ {\rm \mu K\sqrt{s}}$~\cite{PB2instrument}. In the following sections, we use ${\rm NSD}[n]=1 \ {\rm \mu K\sqrt{s}}$ as a fiducial array sensitivity to polarization.

\subsection{Room-temperature instruments}
Here we assume a detection system that has bolometric detectors combined with reflective optics, such as a CMB telescope. Many CMB experiments now use TES bolometers for their small size, which allows a high-density array. Increasing the number of detectors motivates the development of multiplexing techniques. These techniques correlate readout noise with each other among detectors. POLARBEAR-2 and SPT-3G have reflective optics and use the same frequency-domain multiplexing (FDM) readout. See Appendix~\ref{sect:bolometer} for the bolometer principle and FDM readout. AdvancedACT also has a reflective system, but the readout method is time-domain multiplexing (TDM), which causes differences in subdominant terms. Application to BICEP3 requires us drastic modifications of the models because BICEP3 uses refractive optics and TDM readout.

Current CMB experiments aim to precisely measure the CMB polarization pattern. One method of extracting polarization is pair-differences. This takes a difference in signals from two detectors that are sensitive to two orthogonal linear polarizations. This method is simple in concept, but the difference in responsivities of the two detectors induces systematic error. Another way is modulating polarization with a rotating half-wave plate before the light arrives at the detector. This can extract polarization from only one detector and can improve sensitivity at low frequency, but requires special care to optically characterize the half-wave plate and stability of the rotation encoder~\cite{WHWP,Satoru}. To simplify the signal model, we consider excess noise induced in the polarization by instrumental temperature fluctuation in the pair-difference case.

The requirements are set at a certain frequency of interest. The frequency of interest is determined by the scan speed of the telescope and the angular scale of the targeted {\it B}-mode in the case of CMB observation. The scan speed of POLARBEAR is $0.4^\circ /{\rm s}$~\cite{PB_largepatch2019} and the inflationary {\it B}-mode has its maximum amplitude at an approximately 3.6$^\circ$ scale. Therefore the temperature needs to be stable over an interval of 9 seconds. We impose all requirements at double this duration, namely, 50 mHz (20 seconds).

\subsubsection{Readout electronics}
\begin{figure}
\begin{center}
\begin{tabular}{c}
\includegraphics[height=6cm]{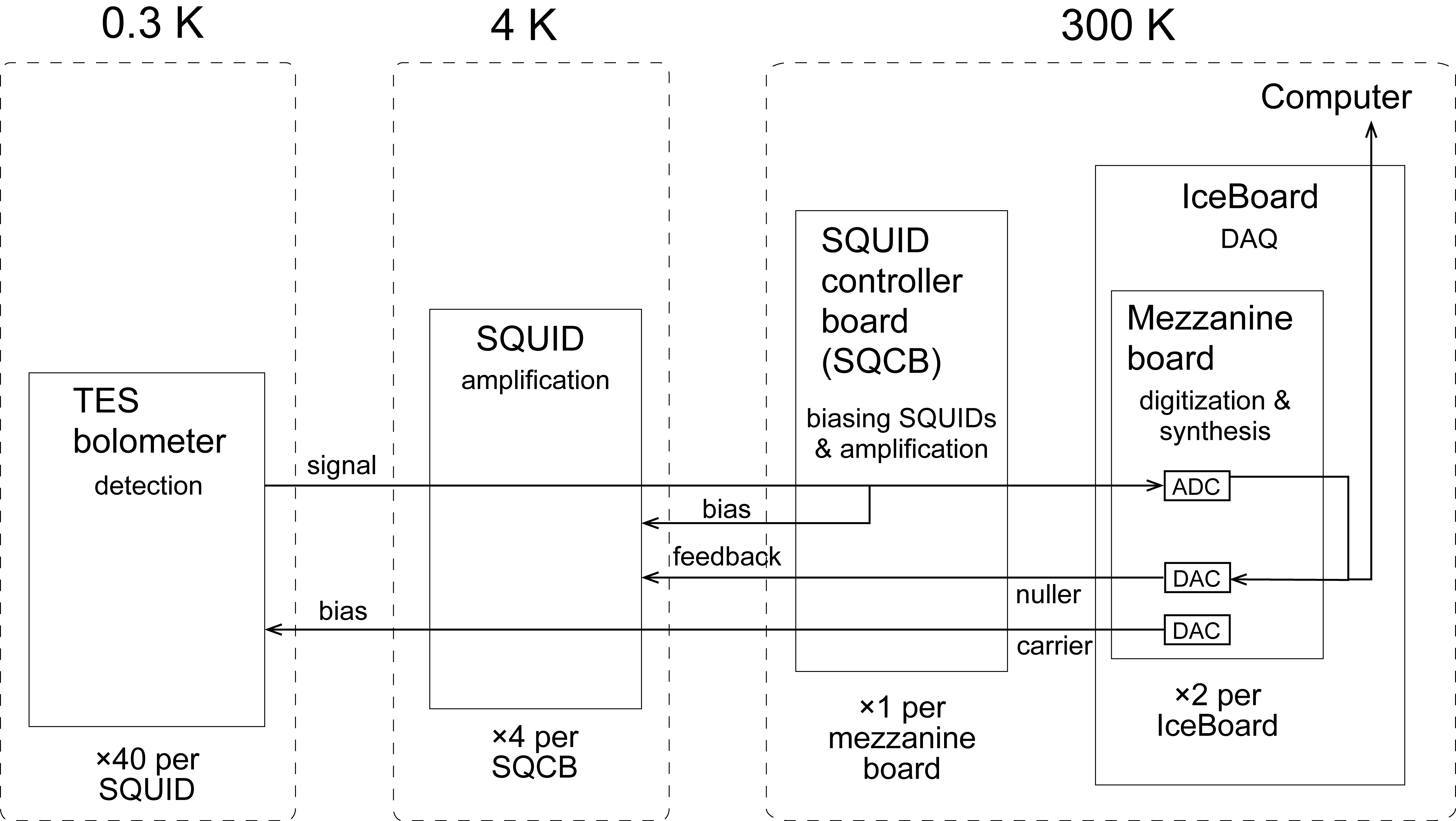}
\end{tabular}
\end{center}
\caption 
{ \label{fig:readout_chains}
Simplified diagram of the readout chain used in POLARBEAR-2.
}
\end{figure}

We consider the temperature coefficient of a readout electronics system based on the system used in POLARBEAR-2 and SPT-3G~\cite{PB2_readout,spt3g_readout}. As shown in Fig.~\ref{fig:readout_chains}, the system consists of IceBoard motherboards, mezzanine boards, and superconducting quantum interference device (SQUID) controller boards (SQCB)~\cite{ice}. A mezzanine board, which is stacked on an IceBoard, communicates with a field-programmable gate array (FPGA) and processes signals to/from the detectors. An SQCB applies bias current to the cooled SQUIDs and amplifies the signal from the SQUIDs. Here we focus on the mezzanine board because the gain fluctuation of an amplifier chain in the SQCB is compensated by digital active nulling (DAN) feedback~\cite{DAN}.

The mezzanine board applies a constant bias voltage to the TESs, reads signals from the SQUIDs, and sends a DAN feedback signal to the TESs to stabilize the SQUIDs at their operating point. These signals fluctuate owing to the temperature dependence of the resistors and ICs in the circuit.

In the pair-difference case, the fluctuation in bias voltage is the major source of the excess polarization noise from the readout electronics. This couples with the non-uniformity in the bias voltages applied to the pair detectors and is induced in the polarization signal. Then the temperature coefficient between the electronics temperature fluctuation and polarization noise, $a_{\rm elec}$, can be expressed as
\begin{equation}
a_{\rm elec}=\delta_{\rm bias}\frac{I_{\rm bias}}{g_0}A_{\rm bias},
\end{equation}
where $\delta_{\rm bias}$ is the fraction of non-uniformity in the bias voltage, $g_0$ is a conversion from the Rayleigh--Jeans temperature to the detector current, $I_{\rm bias}$ is the bias power applied to each TES, and $A_{\rm bias}$ is a fractional coefficient between the circuit temperature and bias voltage. The $\delta_{\rm bias}$ is estimated to be $0.5$\% in POLARBEAR~\cite{PB2017}. The bias power in the temperature unit, $I_{\rm bias}/{g_0}$, is near 20 K for the Chilean observation. Assuming that the total temperature dependence of bias generation circuit is $500$ppm/K~\cite{sqcb_subdominant}, we obtain the required precision of the readout temperature monitor as
\begin{equation}
{\rm NSD\left[ T_{elec}\right]}<\left| \frac{{\rm NSD}[n]}{\displaystyle a_{\rm elec}} \right|\sim 2.0 \ {\rm mK\sqrt{s}}.
\end{equation}

\subsubsection{Mirror}
The polarized emission and reflection from the mirror can be estimated using the formalism of Strozzi \& McDonald (2000) based on the Fresnel law~\cite{strozzi_emissivity}. Their temperature dependence is simply linear under the Rayleigh--Jeans approximation. The dominant polarization components are the polarized emission from the mirror itself and polarization generated from the sky radiation when it is reflected on the mirror surface. The total power of these components is expressed in the Rayleigh--Jeans temperature unit as~\cite{CAPMAP_pol_2005}
\begin{equation}
\label{eq:app_pol_mirror}
P_{\rm mirror,pol}^{\rm RJ}\simeq \displaystyle \sqrt{4\pi \rho \nu \epsilon_0}\frac{\sin^2 \theta}{\cos \theta} \left( T_{\rm mirror}-T_{\rm sky} \right),
\end{equation}
where $\rho$ is the resistivity of the mirror surface, $\nu$ is the frequency of light, $\epsilon_0$ is the permittivity of vacuum, and $\theta$ is the nutation angle between the receiver and the mirror. The quantities $T_{\rm mirror}$ and $T_{\rm sky}$ are the temperatures of the mirror and sky, respectively. Here we assumed that the permittivity of the air is the same as that of vacuum. Note that $\rho$ has a slight temperature dependence.

The temperature dependence of this polarization is derived via
\begin{equation}
\label{eq:app_dependence_pol_mirror}
\begin{array}{lll}
a_{\rm mirror}&\equiv&\displaystyle \frac{dP_{\rm mirror,pol}^{\rm RJ}}{dT_{\rm mirror}} \\
    &=&\displaystyle \sqrt{4\pi \rho \nu \epsilon_0}\frac{\sin^2 \theta}{\cos \theta}\left\{ 1-\frac{1}{2}\frac{d\ln \rho}{dT_{\rm mirror}}\left( T_{\rm mirror}-T_{\rm sky} \right) \right\},
\end{array}
\end{equation}

Here we substitute parameters based on POLARBEAR-2. We fix the mean temperature of the sky and the mirror as $T_{\rm sky}=10$ K~\cite{Hasegawa_coldload} and $T_{\rm mirror}=273$ K. Assuming the mirror made of aluminum, we use $\rho=2.417\times 10^{-8} \ \Omega\cdot {\rm m}$ at 273 K~\cite{Desai_aluminum} and $d\ln \rho/dT_{\rm mirror}\sim 0.004$. The observation frequencies of light are $\nu=85.2$ and $143.9$ GHz~\cite{Kaneko_deploymentPB2A}. Because we are concerned with the upper limits of the temperature coefficients, we use $\nu=143.9$ GHz as the light frequency. The angle between the mirror surface and the detector plane is $\theta=32.5^\circ$\cite{Satoru}. Then we obtain $a_{\rm mirror}\sim 0.032$\%.

When this temperature coefficient is substituted into Eq.(\ref{eq:NSD}), the required precision of the mirror temperature monitor is
\begin{equation}
{\rm NSD\left[ T_{mirror}\right]}<\left| \frac{{\rm NSD}[n]}{a_{\rm mirror}} \right|\sim 3.1 \ {\rm mK\sqrt{s}}.
\end{equation}

For a Chilean telescope, the polarization signal from the mirror is maximally modulated at sunrise and sunset because the temperature drift reaches 20 K on the mirror surface at these times.

The required precision of the temperature monitor is on the order of ${\rm mK\sqrt{s}}$ for both the readout electronics and the mirror. 

We can evaluate the impacts of these temperature fluctuations to the detector array noise using the temperature coefficients and measured temperature noise spectra. The impacts of the room-temperature components are comparable. They are also comparable to those of cold components such as the focal plane. The temperature coefficient of the focal plane is thousands of times larger while the magnitude of temperature fluctuation is on the order of ${\rm \mu K\sqrt{s}}$~\cite{tanabe_phd}.

\section{Design of a room-temperature monitoring system}
\label{sect:design}
\subsection{Design concept}
We realized a system that can monitor temperature fluctuation at 300 K with precision on the ${\rm mK\sqrt{s}}$ level. Instead of facing difficulty in achieving an accuracy of $O(1)$ mK with a room-temperature thermometer, we focused on only tracking relative fluctuation. This motivates us to choose a sensitive sensor even if it has uncertainty in absolute temperature. We combined a low-noise measurement device with thermistors to push down the precision limit determined by measurement noise.

We inserted a scanner to sense multiple thermometers on a large mirror simultaneously. In addition, we ensured that the sampling rate of the temperature monitors was higher than 100 mHz because we are concerned with temperature fluctuation on the 50 mHz scale to study inflation.

We designed two types of room-temperature monitoring systems because the one used for readout electronics needs special consideration for electromagnetic compatibility. The temperature monitoring system for readout electronics is shielded by a Faraday cage so as not to induce excess noise in the readout through its electromagnetic emission.

\subsection{Measurement device}
\label{sect:device}
We constructed a temperature monitoring system with an LTC2983 temperature measurement system~\cite{LTC2983}. The LTC2983 can multiplex up to 10 channels with an optional DC2210 terminal board and is controlled by a Linduino One microcontroller.

We controlled the whole monitoring system with a Raspberry Pi 3B+ single-board computer. The sampling rate was 0.7 Hz. The nominal uncertainty in readout voltage was ${\rm \sim 1 \ \mu V_{rms}}$, corresponding to 0.01 $\Omega$ when the excitation current is 100 ${\rm \mu A}$.

We evaluated the system noise using a metal film resistor with a resistance of 3 ${\rm k\Omega}$ at 298 K~\cite{Vishay_metalfilm}. The nominal temperature coefficient of this resistor was 10 ppm/K, which was 10 times lower than the temperature coefficients of the mirror and the readout electronics. We monitored the resistance of this reference resistor with the LTC2983 system for an hour, and calculated the noise spectrum density (NSD) for every 1000 seconds and averaged the NSDs. Because the internal resistance of the LTC2983 itself has temperature dependence, we monitored the internal resistance using a shorted channel during the measurement and subtracted it from the data. 

\begin{figure}
\begin{center}
\begin{tabular}{c}
\includegraphics[height=6.0cm]{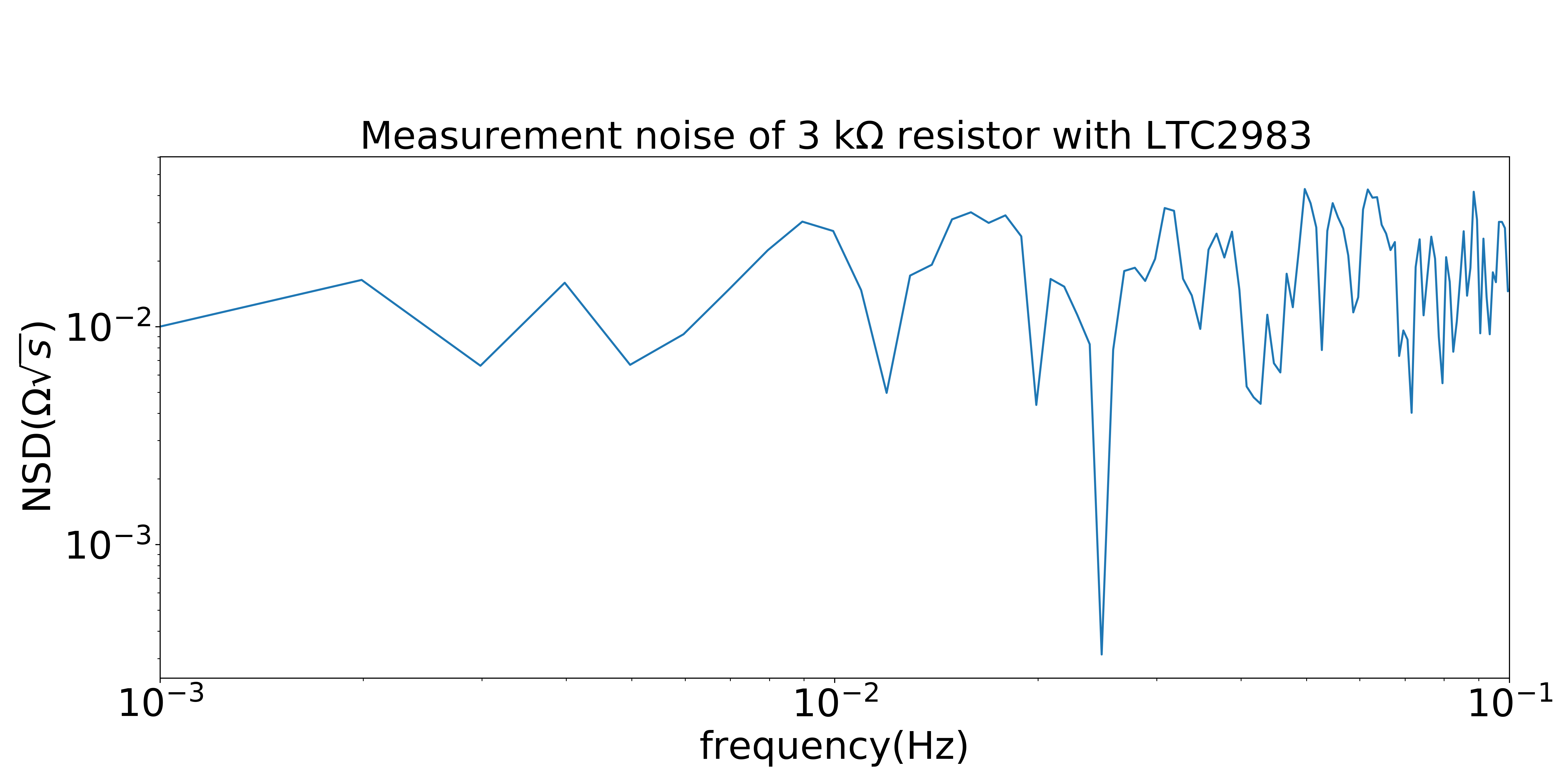}
\end{tabular}
\end{center}
\caption 
{ \label{fig:measurement_noise_ltc2983}
Resistance measurement noise with a temperature measurement system (LTC2983+DC2210). 
}
\end{figure}

As shown in Fig.~\ref{fig:measurement_noise_ltc2983}, the measured noise level around 50mHz was 0.04 ${\rm \Omega \sqrt{s}}$, which corresponds to 0.3 ${\rm mK \sqrt{s}}$ with the sensitivity of our thermistor. This was consistent with the expected noise level based on the nominal readout uncertainty.

For the mirror, we adopted a Keysight 34980A Multifunction Switch/Measure Unit instead of LTC2983. It can multiplex up to 40 channels of two-wire systems connected via an optional 34925A FET multiplexer and a 34925T terminal block, with a 0.2-Hz sampling rate~\cite{keysight34980A}. This feature is suitable for monitoring the temperature of a wide area of the mirror with multiple thermometers. 

We also measured the internal measurement noise of 34980A. The measurement noise at 50mHz was ${\rm 0.06 \ \Omega\sqrt{s}}$, corresponding to ${\rm 0.5 \ mK\sqrt{s}}$ for our thermistor, which satisfies our requirement for monitoring the mirror temperature. This was also consistent with the expected noise level calculated from the nominal measurement uncertainty of the device.

\subsection{Thermometer}
We need to choose a highly sensitive thermometer to track a small fluctuation. The sensor sensitivity can be expressed as the ratio between the change in a sensor observable and the temperature drift.

We chose a thermistor for the thermometer. This is a ceramic resistor with a nearly linear responsivity at room temperature. The thermistor response is expressed in terms of the B-coefficient, which is determined via
\begin{equation}
\label{eq:B_factor}
R=R_0 \exp \left\{ B\left( \frac{1}{T}-\frac{1}{T_0} \right) \right\},
\end{equation}
using two temperatures $T$ and $T_0$ and the corresponding resistances $R$ and $R_0$. 

We tested a PS302J2 thermistor manufactured by Littelfuse Inc.~\cite{thermistor_PS302J2}. Its nominal resistance is $R=3000 \ \Omega$ at $T=298$ K. The width of each sensor is 2.4 mm. The sensor is guaranteed to operate in the temperature range from $-80 \ ^\circ$C to $+150 \ ^\circ$C, which indicates adaptability to ambient instruments in the major part of a year in the South Pole and Chile.

We calibrated it every 5 ${\rm ^\circ C}$ in the range from $-30 \ ^\circ C$ to $+30 \ {\rm ^\circ C}$ and obtained $B=3739$ in this temperature region. This corresponds to a temperature coefficient of $\sim -123 \ \Omega /{\rm K}$ at 298 K. This temperature coefficient and the measurement noise of LTC2983 give us an expected minimum noise level of ${\rm 0.3 \ mK\sqrt{s}}$. 

\begin{figure}
\begin{center}
\begin{tabular}{c}
\includegraphics[height=7cm]{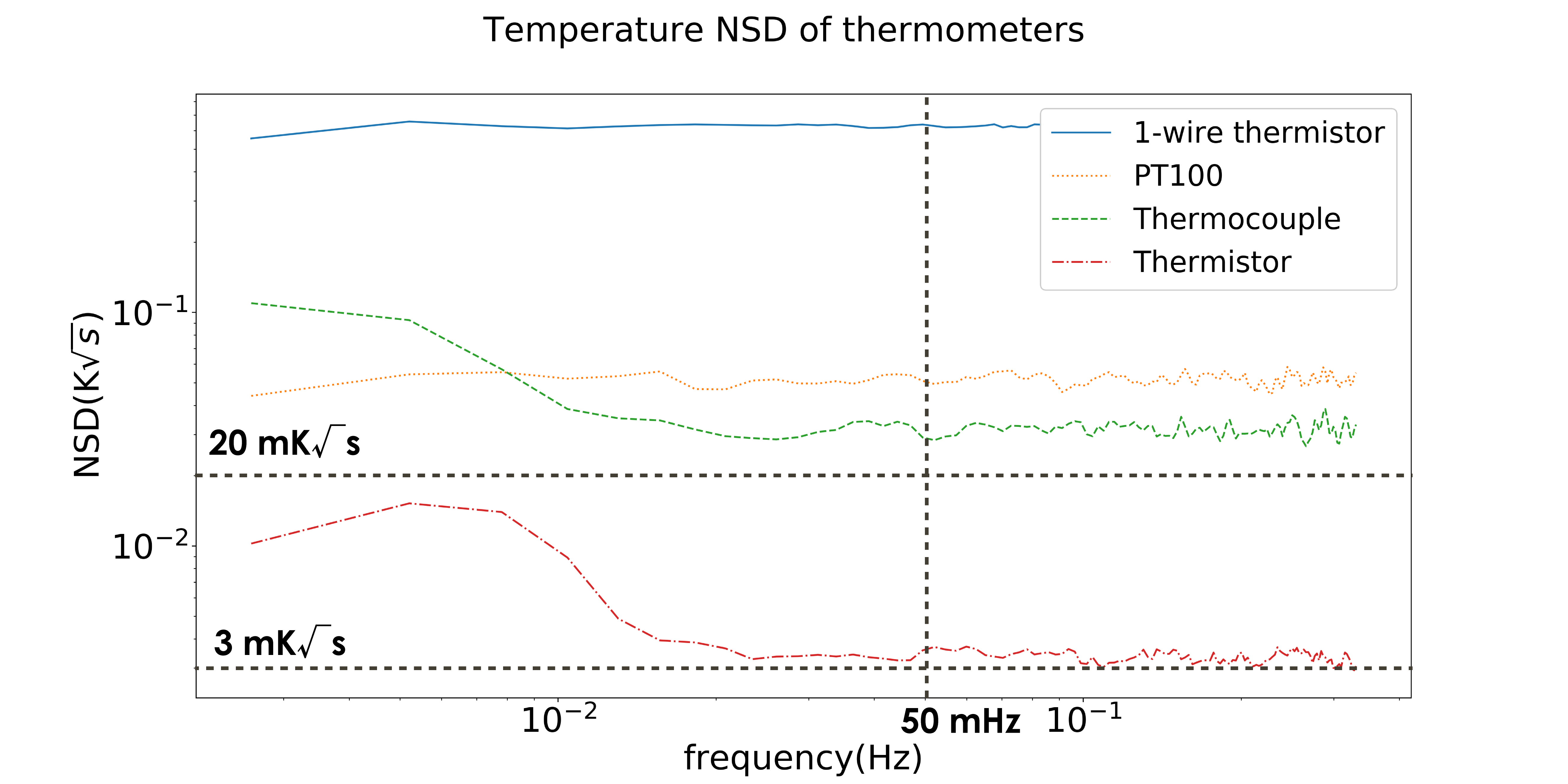}
\end{tabular}
\end{center}
\caption 
{ \label{fig:thermometer_noise}
Temperature measurement noise of thermistor, platinum RTD, T-type thermocouple, and 1-wire temperature sensor. } 
\end{figure}

Fig.~\ref{fig:thermometer_noise} shows a comparison of the noise level with a platinum resistance temperature detector (RTD)~\cite{PTCO_NB-PTCO-011}, thermocouple~\cite{thermocouple_RS}, and 1-wire thermometer~\cite{onewire_DS18B20}. The thermistor had the lowest noise level as expected, which motivated us to choose the thermistor as a temperature sensor.

The noise level can be further pushed down by improving the calibration of the T--R curve of the thermistors and applying a wind protection. As we show in Sec.~\ref{sect:integration}, we finally achieved a noise level of $0.6 \pm 0.2 \ {\rm mK\sqrt{s}}$.

\subsection{Modification for on-site deployment}
\label{sect:integration}
We adopted an LTC2983 as a measurement device for the readout electronics and a Keysight34980A for the mirror. The thermistor was chosen as a thermometer.

\begin{figure}
\begin{center}
\begin{tabular}{c}
\includegraphics[height=7cm]{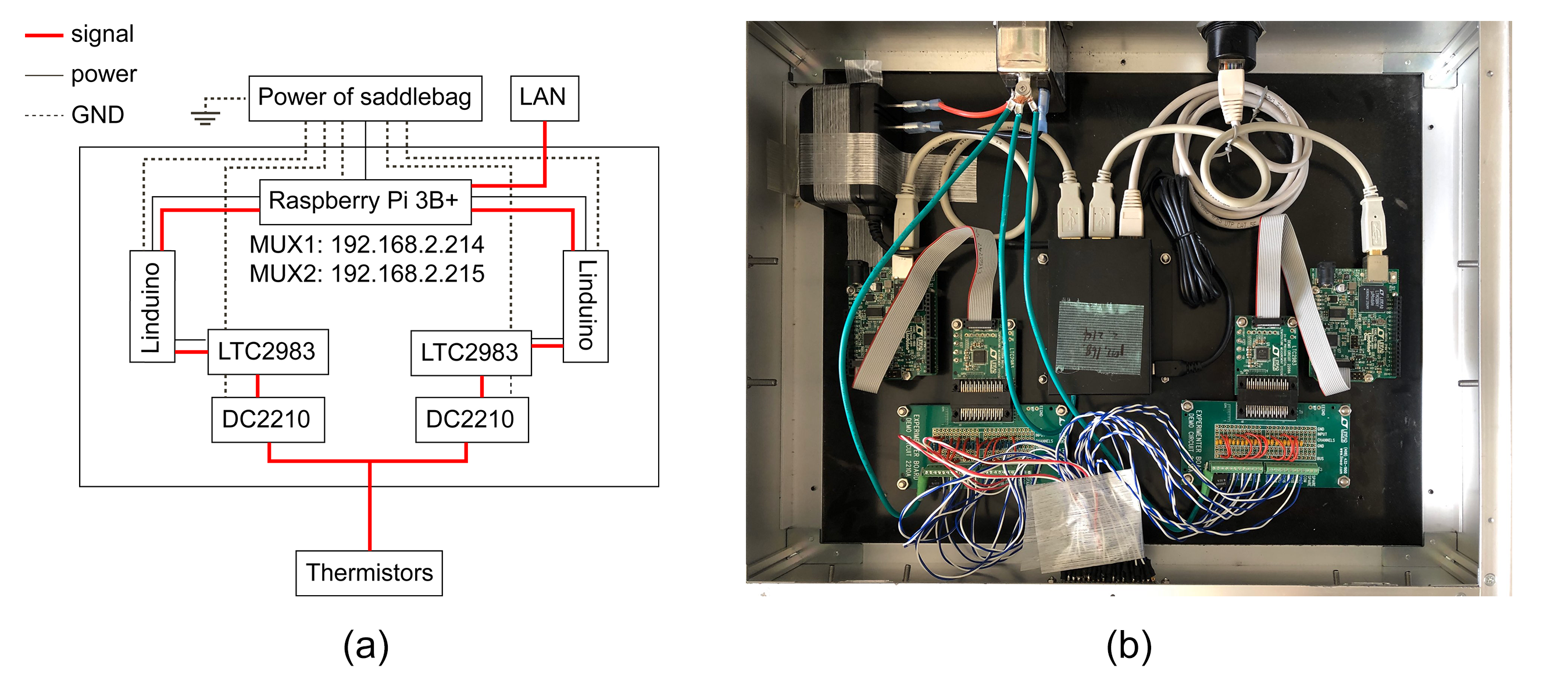}
\end{tabular}
\end{center}
\caption 
{ \label{fig:elec_scheme}
Temperature monitoring system for readout electronics. (a) Schematic. (b) Photograph. The measuring system is located in a metal box for RF shielding. 
} 
\end{figure}

We placed LTC2983 and its peripheral electronics in an aluminum box to shield the electromagnetic emission from themselves. Its schematic and photograph are shown in Fig.~\ref{fig:elec_scheme}. There are two LTC2983 systems in the box. The box chassis is grounded through the power ground.

\begin{figure}
\begin{center}
\begin{tabular}{c}
\includegraphics[height=4cm]{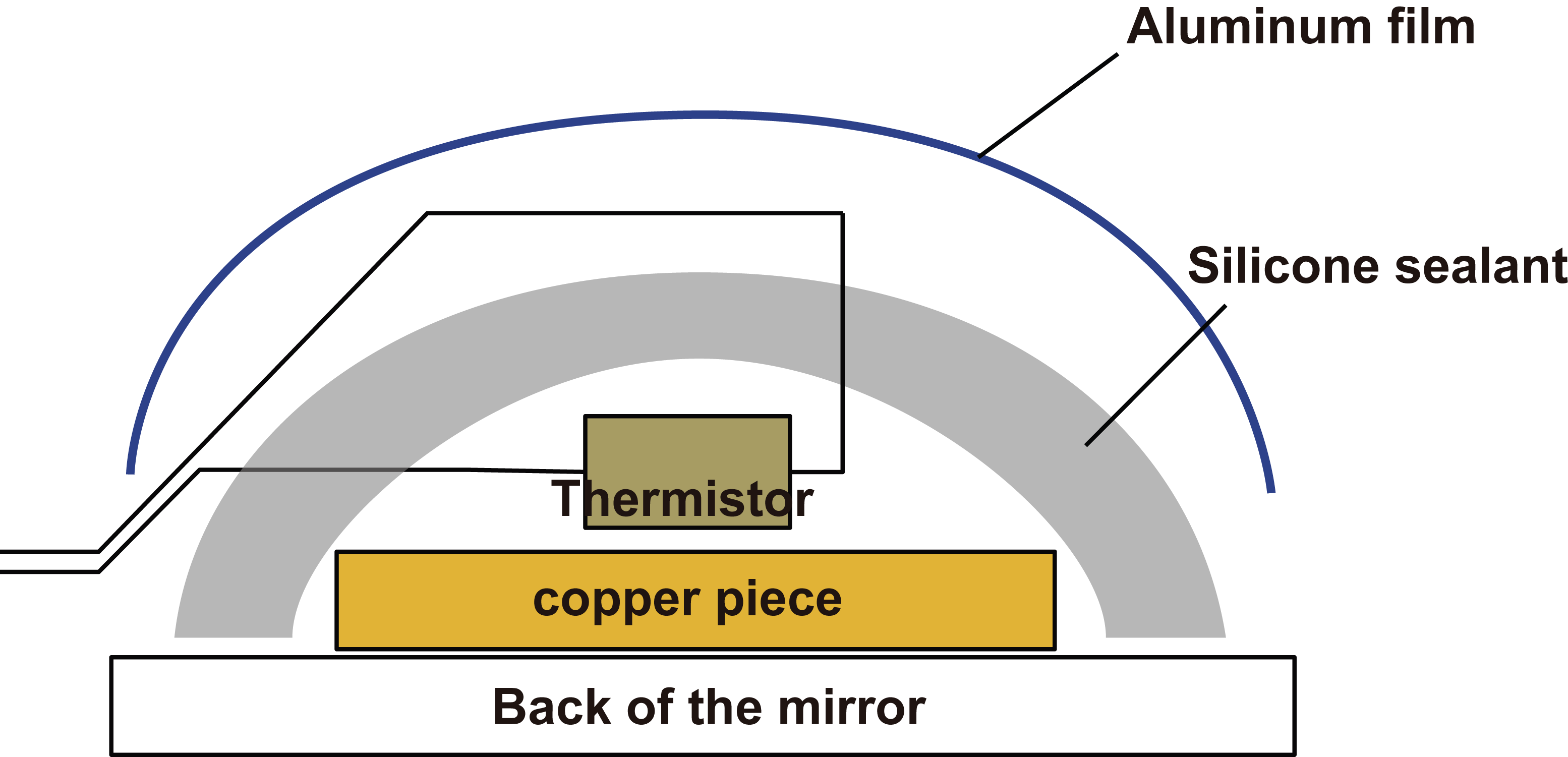}
\end{tabular}
\end{center}
\caption 
{ \label{fig:thermistor_image}
Scheme of the thermistor covered with silicone wind protection and aluminum film.} 
\end{figure}

Figure~\ref{fig:thermistor_image} shows a schematic of wind protection and a sunshade applied to each thermistor. First, we bonded a thermistor to a small and thin copper sheet with an area of 10 mm square and thickness of 1 mm with varnish. Then we covered it with silicone sealant and a thin reflective aluminum film. These applications reduce the temperature fluctuation caused by wind, mitigate the absolute temperature error caused by sunlight, and increase durability in the severe environment of astronomical observatory.

\begin{figure}
\begin{center}
\begin{tabular}{c}
\includegraphics[height=6cm]{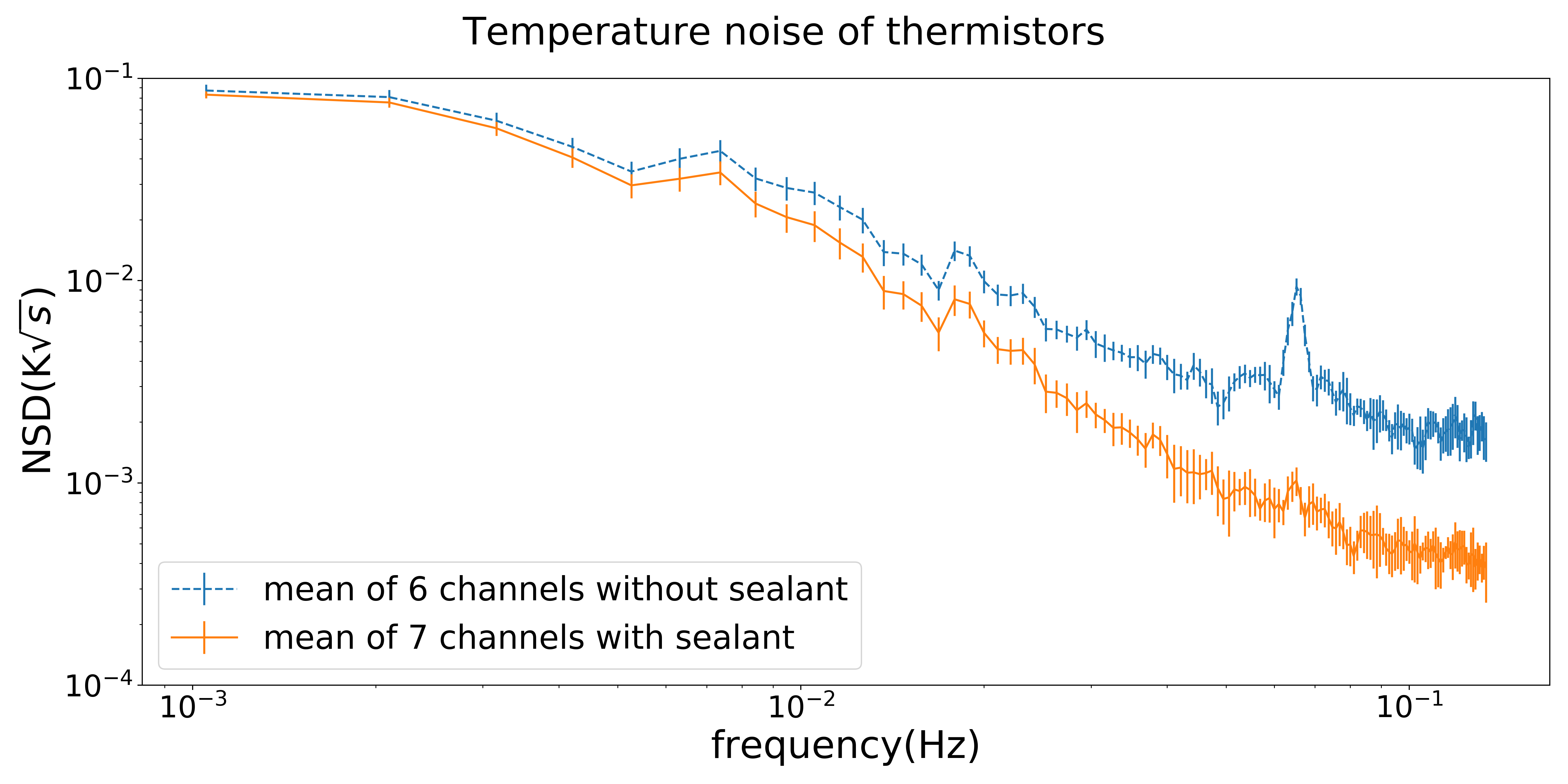}
\end{tabular}
\end{center}
\caption 
{ \label{fig:thermistor_sealant}
Thermal noise of thermistors with- and without- silicone wind protection, measured with 34980A+34925A. Each error bar is the standard deviation of an 850-second window. The blue line is the averaged spectrum of the non-sealant group. The peak around 65 mHz is due to the warm wind blowing in the thermal chamber. The orange line is the averaged spectrum of the sealant group.} 
\end{figure}

We measured the temperature fluctuation of this covered thermistor with the 34980A in the laboratory. A group of thermistors was located in a thermal chamber where hot wind circulates to control temperature. Half of them were covered with silicone sealant and the other half were not. All of them were thermally anchored to the metal floor of the thermal chamber through the copper sheet. We calculated NSDs from 850-second cuts of time-ordered temperature data and obtained ${\rm 2.5\pm 0.4 \ mK\sqrt{s}}$ at 50 mHz for the non-sealant group and ${\rm 0.8\pm 0.2 \ mK\sqrt{s}}$ for the sealant group. The NSDs are shown in Fig.~\ref{fig:thermistor_sealant}. Without active temperature control by the wind, the noise of the sealant group was further reduced to ${\rm 0.6\pm 0.2 \ mK\sqrt{s}}$. This satisfies both requirements for the temperature monitor of the mirror and the readout electronics, with the pair-difference model.

We installed our warm temperature monitoring systems at the POLARBEAR-2 observation site in February 2019. We attached 12 thermistors on the back of each mirror and 2 on each readout crate, and are continuously monitoring the temperature of these warm instruments. 

\begin{figure}
\begin{center}
\begin{tabular}{c}
\includegraphics[height=7cm]{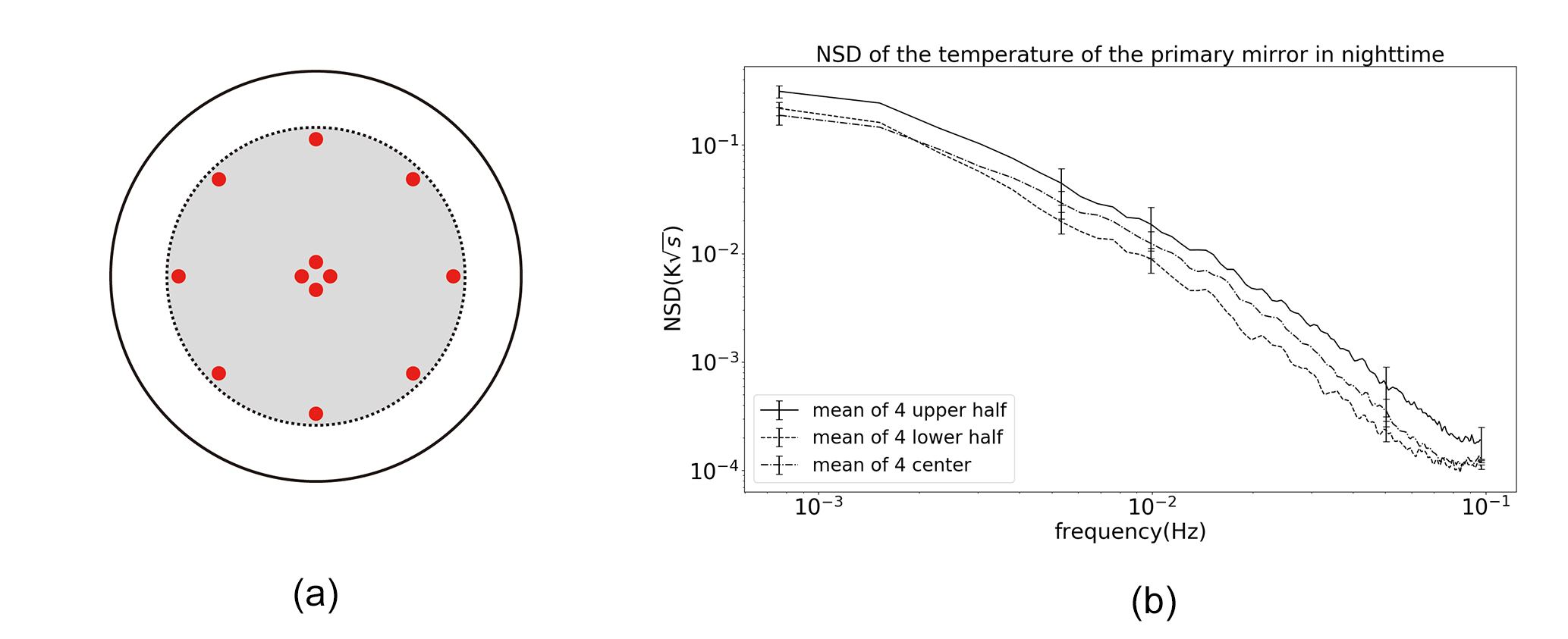}
\end{tabular}
\end{center}
\caption 
{ \label{fig:fluctuation_night}
(a) Position of the 12 thermistors on the back of the primary mirror of POLARBEAR-2. The grey region indicates the actual mirror area. (b) Temperature fluctuation of thermistors on the primary mirror at night. NSDs of 4 thermistors in the upper half, 4 in the lower half, and 4 around the center are averaged. Error bars are the standard deviation of every 4 NSDs.} 
\end{figure}

We measured the noise of the mirror temperature monitor at the observation site. As shown in Fig.~\ref{fig:fluctuation_night}(a), 8 thermistors were radially attached to the outer circumference of the mirror and 4 were around the center. We measured the temperature of the thermistors from 2 a.m. to 8 a.m. (sunrise) and calculated NSDs from 1000-second cuts. Figure~\ref{fig:fluctuation_night}(b) shows averaged NSDs of 4 thermistors in the upper half including the left one viewed from the front, 4 in the lower half including the right one, and 4 around the center. All thermistors have temperature fluctuation below 3 ${\rm mK\sqrt{s}}$, indicating that they can monitor the temperature with sufficiently low measurement noise at the site. The variation of the fluctuation on the mirror could be induced by different conditions of being exposed to wind.

\section{Discussion}
\label{sect:discussion}
We should consider the temperature distribution of the target. For example, temperature gradient between the back and front of a mirror, and that on the front or back surface of a mirror cause errors in measuring absolute temperature and relative temperature fluctuation. 

We suggest reconstructing the temperature gradient on a surface using multiple thermometers and subtracting the temperature fluctuation of a beam-projected area multiplied by the temperature coefficient for every detector. We can compensate for the temperature difference between the back and front of the mirror by modeling the thermal conduction in the material and convection caused by the air.

Another solution for treating a front-to-back temperature gradient is using thermometers and wires thinner than the wavelength of the target light, and directly attaching them to the front surface of the mirror. This is achievable for observing radiation with a wavelength of over a millimeter. Embedding sensitive thermometers in a circuit board is also effective for correcting the temperature gradient around the readout electronics.

Our system for the mirror is designed to read many thermometers to monitor the whole area of a large mirror. We tested 13 thermistors in this study. Our measurement device for the mirror, a Keysight 34980A, can have its multiplexing index increased to 320 by adding 34925A multiplexers. The sampling rate, however, decreases when the number of thermometers increases. 

We estimated the required precision of the temperature monitor was to be on the order of millikelvin, but this can be loosened by configuring observation. For example, a rotating half-wave plate can reduce the dependence on temperature fluctuation of the readout electronics because it extracts the polarization without uncertainty in the relative responsivity of detector pairs.

\section{Summary}
\label{sect:summary}
We developed a system that can measure the temperature of room-temperature instruments of a radio-astronomical telescope with high precision. We made one system for the mirror and the other for the readout electronics. Both of our monitoring systems consist of a low-noise measurement device, a multiplexer, and thermistors. All components were commercial products, and the total cost was approximately 4,000 USD. We achieved a noise spectrum density below ${\rm 1 \ mK\sqrt{s}}$ per channel on a 20-second scale in measurements at 300 K in the laboratory.

Both systems use a thermistor as a thermometer. This is the key to achieving high precision owing to its high sensitivity and low cost. Because the concept is tracking relative temperature fluctuation rather than measuring the absolute value of temperature, we do not precisely calibrate the absolute temperature scale of the thermometers. This enabled us to adopt a thermistor in spite of its large absolute inaccuracy. The configuration of the thermistor attachment was designed to block wind and direct sunlight, and to maintain a fast response.

Our temperature monitoring system is applicable to a ground-based experiment with a detector array with a noise level of $O(1) \ {\rm \mu {\rm K} \sqrt{s}}$. This system will be an essential tool for observing subtle and continuous signals, exemplified by the CMB. 

\section{Acknowledgement}
\label{sect:acknowledgement}
This work was supported by JSPS KAKENHI grant numbers JP14J01662, JP18H01240, JP18H04362, JP18J02133, and the SOKENDAI Short-term Research Abroad \& Long-term Internship Program. This research used resources of the Central Computing System, owned and operated by the Computing Research Center at KEK. We appreciate Dr. Neil Goeckner-Wald, who provided the basic idea of low-frequency noise from instrumental temperature. Dr. Matt Dobbs, Dr. Joshua Montgomery, Dr. Jean-François Cliche, and Dr. Tijmen de Haan provided detailed information on the temperature dependence of the readout circuit. Dr. Grant Teply, Dr. Nils Halverson, Dr. Brad Benson and Dr. Kevin Crowley informed us about temperature monitoring systems in third-generation CMB experiments. Additionally, Dr. Akito Kusaka, Dr. Aritoki Suzuki, Dr. Yuji Chinone and Dr. John Groh gave us much advice on installing the temperature measurement systems. Dr. Andrew May gave us insight in identifying noise sources by their spectra. Mark Kurban from Edanz Group\cite{edanz} edited a draft of this paper.

\appendix
\section{Bolometer system in POLARBEAR-2}
\label{sect:bolometer}
The TES bolometer is a high-resolution energy sensor operated in the intermediate state between normal conduction and superconductivity, so called the ``transition edge'' state. It is kept at this transition edge by applying bias power to balance cooling and optical power. When a photon comes into the heat absorber of the detector, the photon energy raises the temperature and is detected as a sharp decrease in the bias current. This indicates that fluctuation in the applied bias power directly affects the detector signal. The bias voltage amplitude is typically 2 $\mu$V for POLARBEAR-2, which is determined to be comparable to the optical load when the resistance of a TES is $\sim1 \ \Omega$.

Increasing the number of TES bolometers motivates development of multiplexing techniques. This is required for reducing the number of readout lines which make the wiring complicated and transmit thermal load into the cryostat. In the POLARBEAR-2 case, every 40 TES bolometers are read out by one line using digital frequency-domain multiplexing (DfMUX)~\cite{DfMUX}. This method assigns a different resonant frequency to each TES bolometer by connecting it to an LC circuit. Signals from TES bolometers are modulated by the resonant frequencies, accumulated on a readout line, amplified by superconducting quantum interference devices (SQUIDs) on a 4 K stage, and finally demodulated by electronics at the room temperature (Fig.~\ref{fig:readout_chains}). The gain fluctuation of this readout chain makes the signal drift.

\bibliography{reference}

\listoffigures
\listoftables

\end{spacing}

\end{document}